\documentclass[aps,twocolumn,showpacs,preprintnumbers,nofootinbib,prd,superscriptaddress,groupedaddress,10pt]{revtex4-1}

\makeatletter
\def\l@subsubsection#1#2{}
\def\l@subsubsubsection#1#2{}
\makeatother

\usepackage{amsfonts}
\usepackage{graphicx,amssymb,amsmath,amsthm,epsfig,epsf}
\usepackage[usenames]{color}
\usepackage{epstopdf}

\usepackage{aas_macros}
\usepackage{bm}
\usepackage{dcolumn}
\usepackage[latin1]{inputenc}
\usepackage{latexsym}
\usepackage{rotating}
\usepackage{longtable}

\usepackage{enumerate}
\usepackage{tensor,multirow}
\usepackage{url}
\usepackage[linktocpage]{hyperref}
\usepackage{diagbox}
\usepackage{colortbl}
\usepackage{tikz}
\usetikzlibrary{patterns}
\usepackage{tabularx}

\def\nn{\nonumber}

\def\be{\begin{equation}}
\def\ee{\end{equation}}
\newcommand{\beq}{\begin{eqnarray}}
\newcommand{\eeq}{\end{eqnarray}}

\newcommand\tikzmark[2][]{
  \tikz[remember picture,inner sep=\tabcolsep,outer sep=0,baseline=(#1.base),align=left]{\node[minimum width=\hsize](#1){$#2$};}
}
\def\ba{\begin{align}}
\def\ea{\end{align}}

\newcolumntype{Y}{>{\centering\arraybackslash}X}

\newlength{\flexcheckerboardsize}

	\newcommand{\defineflexcheckerboard}[5]{
    \setlength{\flexcheckerboardsize}{#2}
    \pgfdeclarepatterninherentlycolored{#1}
        {\pgfpointorigin}{\pgfqpoint{2\flexcheckerboardsize}    
        {2\flexcheckerboardsize}}
        {\pgfqpoint{2\flexcheckerboardsize}
        {2\flexcheckerboardsize}}%
        {
            \pgfsetfillcolor{#4}
            \pgfpathrectangle{\pgfpointorigin}{
            \pgfqpoint{2.1\flexcheckerboardsize}    
                {2.1\flexcheckerboardsize}}
          \pgfusepath{fill}
          \pgfsetfillcolor{#3}
          \pgfpathrectangle{\pgfpointorigin}
            {\pgfqpoint{\flexcheckerboardsize}
            {\flexcheckerboardsize}}
          \pgfpathrectangle{\pgfqpoint{\flexcheckerboardsize}
            {\flexcheckerboardsize}}
            {\pgfqpoint{\flexcheckerboardsize}
            {\flexcheckerboardsize}}
            \pgfusepath{fill}
        }
}

\defineflexcheckerboard{flexcheckerboard_bw}{.5mm}{black}{white}{0}
\defineflexcheckerboard{flexcheckerboard_redblue}{2mm}{red}{blue}{0.1}
\defineflexcheckerboard{flexcheckerboard_greenorange}{10mm}{green}{orange}{0}
\defineflexcheckerboard{flexcheckerboard_bluecyan}{.2mm}{cyan}{blue}{0}

\begin{document}
\title{Exotic compact objects with soft hair}

\author{
Guilherme Raposo,$^{1}$
Paolo Pani,$^{1}$
Roberto Emparan$^{2,3}$
}
\affiliation{$^{1}$ Dipartimento di Fisica, ``Sapienza'' Universit\`a di Roma \& Sezione INFN Roma1, Piazzale Aldo Moro 5, 00185, Roma, Italy}
\affiliation{$^{2}$ Instituci\'o Catalana de Recerca i Estudis Avan\c cats (ICREA), Passeig Llu\'{\i}s Companys 23, E-08010 Barcelona, Spain}
\affiliation{$^{3}$ Departament de F{\'\i}sica Qu\`antica i Astrof\'{\i}sica, Institut de Ci\`encies del Cosmos, Universitat de
Barcelona, Mart\'{\i} i Franqu\`es 1, E-08028 Barcelona, Spain}

\begin{abstract}
Motivated by the lack of a general parametrization for exotic compact objects, 
we construct a class of perturbative solutions valid for small (but 
otherwise generic) multipolar deviations from a Schwarzschild 
metric in general relativity. 
We introduce two classes of exotic compact objects, with ``soft'' and 
``hard'' hair, for which the curvature at the surface is respectively 
comparable to or much larger than that at the corresponding black-hole horizon.
We extend the Hartle-Thorne formalism to relax the assumption of equatorial 
symmetry and to include deformations induced by multipole moments higher than 
the spin, thus  constructing the most general, axisymmetric quasi-Schwarzschild 
solution to Einstein's vacuum equations. 
We explicitly construct several particular solutions of objects with soft 
hair, which might be useful for tests of quasi-black-hole metrics, and also to 
study deformed neutron stars.
We show that the more compact a soft exotic object is, the less hairy it will be.
All its multipole moments can approach their 
corresponding Kerr values only in two ways as their compactness increases: either 
logarithmically (or faster) if the moments are spin-induced, or linearly (or faster) 
otherwise.
Our results suggest that it is challenging (but possibly 
feasible with next-generation gravitational-wave detectors) to distinguish Kerr black 
holes from a large class of ultracompact exotic objects on the basis of their different 
multipolar structure.
\end{abstract}

\maketitle


\section{Introduction}
%
As a by-product of the black-hole (BH) uniqueness and no-hair
theorems~\cite{Carter71,Hawking:1973uf} (see also~\cite{Heusler:1998ua,Chrusciel:2012jk,Robinson}),
the multipole moments of any stationary BH in isolation can be written
as~\cite{Hansen:1974zz},
\begin{equation}
 \mathcal{M}_\ell^{\rm BH}+i \mathcal{S}_\ell^{\rm BH}  
 =\mathcal{M}^{\ell+1}\left(i\chi\right)^\ell\,, \label{nohair}
\end{equation}
where $\mathcal{M}_\ell$ ($\mathcal{S}_\ell$) are the Geroch-Hansen mass (current) 
multipole moments~\cite{Geroch:1970cd,Hansen:1974zz}, the suffix ``BH'' refers to the Kerr 
metric, and
\begin{equation}
\chi\equiv\frac{\mathcal{S}_1}{\mathcal{M}_0^2}
\end{equation}
is the dimensionless spin. Equation~\eqref{nohair} implies that $\mathcal{M}_\ell^{\rm 
BH}$ ($\mathcal{S}_\ell^{\rm BH}$) 
vanish when $\ell$ is odd (even), and that all moments with $\ell\geq2$ can be written 
only in terms of the mass $\mathcal{M}_0=\mathcal{M}$ and angular momentum 
$\mathcal{S}_1=\mathcal{J}$ (or, equivalently, $\chi$) of the BH. Therefore, any 
independent measurement of three multipole moments (e.g. the mass, the spin and the mass 
quadrupole
$\mathcal{M}_2$) provides a null-hypothesis test of the Kerr metric and, in turn, it 
might serve as a genuine strong-gravity confirmation of
general 
relativity~\cite{Psaltis:2008bb,Gair:2012nm,Yunes:2013dva,Berti:2015itd,
Cardoso:2016ryw,Barack:2018yly}.

Motivated by some scenarios inspired by semiclassical and quantum gravity which 
predict exotic ultracompact objects without a 
horizon (see, 
e.g.,~\cite{Mazur:2004fk,Mathur:2005zp,Mathur:2008nj,Barcelo:2015noa, 
Danielsson:2017riq,Berthiere:2017tms}) or new physics at the horizon scale~ 
(see~\cite{Cardoso:2017cqb,Cardoso:2017njb,Carballo-Rubio:2018jzw} for some 
overview), the aim of 
this paper is to identify some generic features in the multipolar structure of 
exotic compact objects~(ECOs) and to construct explicit quasi-BH solutions to Einstein's 
equations in vacuum. 

The vacuum region outside a spinning object is not generically described by the Kerr 
geometry, due to the absence of an analog to the Birkhoff's theorem in axisymmetry. Thus, 
the multipole moments of an axisymmetric ECO~(spinning or not) will generically 
satisfy relations of the form
\begin{eqnarray}
  \mathcal{M}_\ell^{\rm ECO} &=& \mathcal{M}_\ell^{\rm BH} +\delta \mathcal{M}_\ell \,, 
\label{mmECOM}\\
  \mathcal{S}_\ell^{\rm ECO} &=& \mathcal{S}_\ell^{\rm BH} +\delta 
\mathcal{S}_\ell \,,\label{mmECOS}
\end{eqnarray}
where $\delta \mathcal{M}_\ell$ and $\delta 
\mathcal{S}_\ell$ are model-dependent corrections, whose precise value can be obtained by 
matching the metric describing the interior of the object to that of the exterior.
We shall assume that matter fields are confined in the interior and that the exterior is 
governed by Einstein's equations. This is a nontrivial assumption on the physics of the 
ECO, namely that the exotism\footnote{Although our main motivation is to study ECO 
spacetimes, we shall construct a general perturbative solution of vacuum 
Einstein's equations, which might be useful also to study deformed neutron stars within 
general relativity.} resides entirely in the matter and not in the gravity sector. It 
allows us to perform an analysis of great generality independently of the details of the 
exotic interior: different models of ECOs are defined by the 
boundary conditions at the object's surface, which uniquely define $\delta 
\mathcal{M}_\ell$ and $\delta\mathcal{S}_\ell$.

We introduce two classes of ECO models:
\begin{itemize} 
 \item \emph{``soft'' ECOs}: for which the boundary conditions are such that the 
curvature at the surface is comparable to that at the horizon of the corresponding BH, 
i.e. ${\cal K}^{1/2}\sim 1/{\cal M}^2$ (we adopt $G=c=1$ units throughout and use the 
Kretschmann scalar ${\cal K}$ as a measure of the compactness). This corresponds to 
models in which there is no further length scale in the exterior other than ${\cal M}$ 
or in which the new scale is parametrically close to ${\cal M}$. An example of the latter 
case is the ``scrambling time''~\cite{Hayden:2007cs,Sekino:2008he}, $\sim -{\cal 
M}\log({\cal L}/{\cal M})\gtrsim {\cal M}$, where even if there is a new length scale 
${\cal L}\ll{\cal M}$, its effect on the curvature at the surface is logarithmically 
suppressed.
In other words, in these models the near-surface geometry is similar to that of a BH
and smoothly approaches the horizon in the BH limit~(hence their ``softness'').
 \item \emph{``hard'' ECOs}: for which the curvature at the surface can be much 
larger than that at the horizon of the corresponding BH, presumably because the 
underlying theory involves a new length scale, ${\cal L}$, such that ${\cal 
L}\ll {\cal 
M}$, so the ECO can support large curvatures on its surface without collapsing. The 
curvature in this case typically behaves like a power of ${\cal M}/{\cal L}$, hence ${\cal 
K}^{1/2}\gg1/{\cal M}^2$. In other words, in these models 
high-energy effects drastically modify the near-surface geometry (hence their 
``hardness'').
\end{itemize}

In this work with focus only on soft ECO solutions (which we refer to 
also as ``ECOs with soft hair''\footnote{Note that the name is unrelated to the proposal 
of BHs with soft hair produced by supertranslation symmetries~\cite{Hawking:2016msc}. 
We nevertheless find the name appropriate since, for high compactness, soft ECOs 
must have vanishingly small multipolar deviations (i.e., ``soft hair'') relative to a Kerr 
BH, as we shall show.}); an analysis of certain hard ECO models 
will appear elsewhere~\cite{inprep}. 

Soft ECOs are characterized by multipolar hair that do not by itself require physics 
on a short-distance scale, but this does not mean that such ECOs are possible in the 
absence of any very high-energy physics. Indeed, an ECO with a surface just above the 
BH limit requires large internal stresses in order to prevent its collapse, even if 
the exterior is exactly the Schwarzschild or Kerr geometry.  Our notion of soft ECOs 
refers to the scale of the physics that is implied by the existence of multipolar hair. 
This is a feature distinguishable purely from the ECO exterior, and therefore, to the 
extent that this exterior is governed by Einstein's theory, the characterization of the 
softness of an ECO is very much model-independent.
Likewise, the exterior of hard ECOs might be described 
by soft ECO solutions far from the surface, where the curvature is 
perturbatively close to that of a BH.

In this paper, we shall construct perturbative solutions obtained by 
solving the vacuum Einstein equations order by order in a small multipole moment 
expansions.
We shall classify the solutions in terms of the type of independent multipole moments that 
they possess to leading order. In this scheme, each solution can possess an infinite 
tower of multipole moments which are sourced by the leading-order ones. For 
example, a solution possessing only ${\cal S}_1={\cal J}$ to leading order will contain an 
infinite number of multipole moments at higher orders. We shall refer to the latter as 
\emph{spin-induced} moments since they vanish as ${\cal J}\to0$. On the other hand, we 
shall refer to the multipole moments that remain nonzero as ${\cal J}\to0$ as 
\emph{nonspin-induced} moments.

We shall provide evidence for the 
following conjecture. In the BH limit, the deviations 
from the Kerr multipole moments (with $\ell\geq2$) vanish as
\begin{eqnarray}
 \frac{\delta \mathcal{M}_{\ell}}{\mathcal{M}^{\ell+1}} &\to& 
a_\ell\frac{\chi^{\ell}}{\log\delta}+b_\ell\, \delta +...\,,\label{conjectureM}\\
 \frac{\delta \mathcal{S}_{\ell}}{\mathcal{M}^{\ell+1}} &\to& 
c_\ell\frac{\chi^{\ell}}{\log\delta}+d_\ell\, \delta +...\,, \label{conjectureS}
\end{eqnarray}
or \emph{faster}. Here $a_\ell$, $b_\ell$, $c_\ell$, and $d_\ell$ are numbers of order 
unity or smaller, the ellipsis stand for terms which are subleading in 
our expansion, whereas $\delta\ll1$ is a dimensionless number that can be expressed in 
terms of coordinate-independent geometrical quantities and measures 
the compactness of the object in a way to be specified below. The BH limit 
corresponds to $\delta\to0$. The coefficients $a_\ell$ and $c_\ell$ are related to the 
spin-induced contribution to the multipole moments; selection and $\mathbb{Z}_2$ 
rules presented below imply that $a_\ell$ and $c_\ell$ be identically zero when 
$\ell$ is odd and even, respectively. On the other hand, the coefficients $b_\ell$ and 
$d_\ell$ are related to the nonspin-induced contributions up to order $\ell$.
Note that the selection rules imposed by the equatorial symmetry 
of the Kerr solution do not necessarily apply to ECOs, so 
$\mathcal{M}_\ell^{\rm ECO}$ ($\mathcal{S}_\ell^{\rm ECO}$) can be nonzero also 
when $\ell$ is odd (even) and the solution can break the equatorial symmetry 
through the terms proportional to $b_\ell$ and $d_\ell$ in 
Eqs.~\eqref{conjectureM} and \eqref{conjectureS}.

In other words, in this perturbative scheme the deviations from 
the Kerr multipole moments must die sufficiently fast as the compactness of the 
object approaches that of a BH, or otherwise the curvature at the surface will grow and 
the models cannot classify as soft ECOs. 
As the ECO approach the BH limit, spin-induced moments are less strongly suppressed
than other moments and are therefore easier to detect.

In the rest of this work we quantify the above statements by constructing a 
perturbative expansion which is valid for generic axisymmetric\footnote{In the case of spinning geometries, we 
assume that the angular momentum is aligned with the axis of symmetry. Hence, 
the solution is stationary and, in particular, there is no precession nor gravitational-wave emission.} 
solutions to vacuum Einstein's equations with small multipole 
moments.
Our perturbative scheme includes the Hartle-Thorne 
solution~\cite{Hartle:1967he,1968ApJ...153..807H} as a particular 
case, and it extends it by including generic small corrections that 
break the equatorial symmetry and deformations that are induced by 
multipole moments other than the spin.

\begin{figure*}[th]
\centering
\includegraphics[width=1\textwidth]{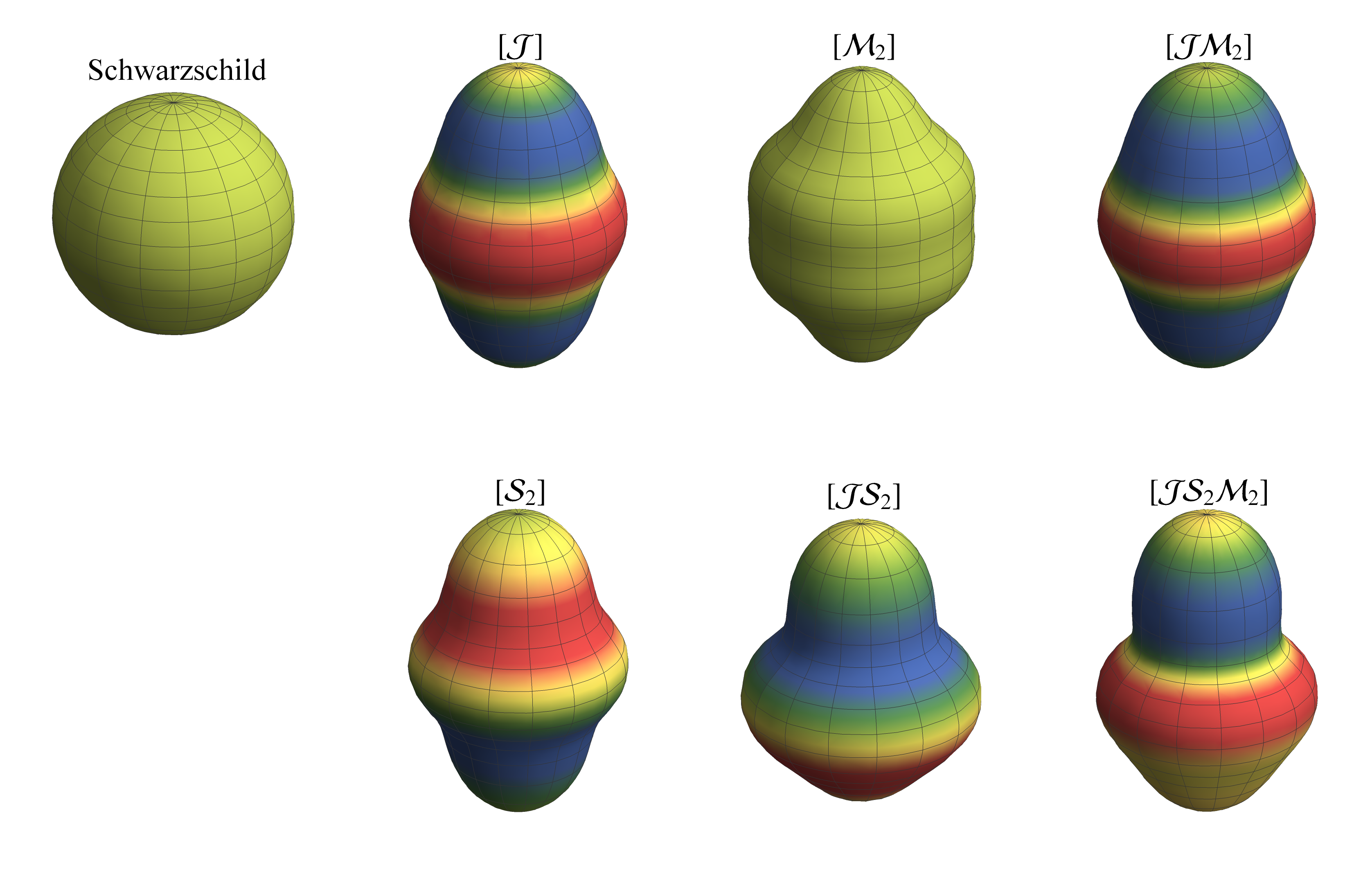}
\caption{Embedding diagrams for some representative soft ECO solutions presented in the 
text. We show the embedding of $g_{\theta\theta}d\Omega^2=r^2(1+\sum_{n,\ell}\epsilon^n 
K^{n\ell}P_\ell)d\Omega^2$ for surfaces of constant $r_0$ and $t$. The colors are weighted 
according to $g_{t\varphi}$ to represent the current multipole moments. The first row 
contains equatorially-symmetric solutions, as evident by the symmetry between 
the North and South hemispheres in each diagram. The second line contains nonequatorially 
symmetric solutions: the differences between the North and South 
hemispheres are either in the colors (for ${\cal S}_\ell$ with even $\ell$), or in the 
shape (for ${\cal M}_\ell$ with odd $\ell$). The shape of the $\left[{\cal 
S}_2\right]$-solution is equatorially symmetric since it does not contain mass 
multipole moments of odd order. For illustrative purposes the 
multipolar deviations $\delta{\cal M}_\ell$ and $\delta{\cal S}_\ell$ are chosen 
artificially large to magnify the effect.}\label{fig:embedding}
\end{figure*}

Several interesting particular solutions are explicitly discussed below and their metric 
is publicly available in closed form in an online repository containing 
supplemental {\scshape Mathematica}\textsuperscript{\textregistered} 
notebooks~\cite{webpage}. As an anticipation of our main result, Fig.~\ref{fig:embedding} 
presents the embedding diagrams for some representative solutions.

Henceforth we adopt the Geroch-Hansen definition of the multipole 
moments~\cite{Geroch:1970cd,Hansen:1974zz}; the latter are 
equivalent~\cite{Gursel} to the multipole moments defined by 
Thorne~\cite{Thorne:1980ru} using asymptotically mass-centered Cartesian 
coordinates.

\section{ECOs with soft hair: a general small-multipole expansion} 
\label{sec:smallmoments}

In this section we implement our method for characterizing the soft hair of an ECO.
Our assumptions are that
\begin{enumerate}
\item the exterior of the ECO is governed by Einstein's vacuum equations,
\item its geometry deviates by a small amount from the Schwarzschild metric.
\end{enumerate}
We consider generic stationary axisymmetric deviations which include not only the soft 
hair but also angular momentum, so these ECOs can rotate slowly. By inserting these 
perturbations in the vacuum Einstein equations and decomposing them into spherical 
harmonics of degree $\ell$, we obtain a set of ordinary differential equations for each 
multipolar mode, which we solve analytically. After requiring asymptotic flatness at 
infinity, we must still give boundary conditions at the ECO surface. These can be 
completely specified, at each perturbative order, by a pair of constants $M_\ell$ and 
$S_\ell$ for each $\ell\geq2$. These constants are directly connected to the physical mass 
and current multipoles, $\mathcal{M}_{\ell}$ and $\mathcal{S}_{\ell}$, and therefore 
provide a general, physical parametrization of the soft hair.

\subsection{Setup}
We consider a metric of the form
\begin{equation}\label{eq:pertmetric}
g_{\mu\nu}=g_{\mu\nu}^{(0)}+\sum_{n=1}^\infty \epsilon^n h^{(n)}_{\mu\nu}\,,
\end{equation}
where $g_{\mu\nu}^{(0)}$ is the background Schwarzschild metric in Schwarzschild 
coordinates, $\epsilon$ is a small book-keeping parameter, and $h^{(n)}_{\mu\nu}$ is the 
perturbation entering at order ${\cal O}(\epsilon^n)$. Note that, although we 
introduced a single book-keeping parameter $\epsilon$, there might exist several physical 
expansion parameters, one for each multipole moment introduced at the leading 
order\footnote{For example, if the spin ${\cal J}$ and the mass quadrupole 
${\cal M}_2$ are present at the leading order, the physical expansion parameters are 
the dimensionless quantities $\epsilon {\cal J}/{\cal M}^2$ and $\epsilon 
{\cal M}_2/{\cal M}^3$; the latter are independent from one another.}.

We focus on stationary and axisymmetric perturbations, expand them in a complete 
basis of spherical harmonics (reducing to Legendre polynomials $P_\ell(\cos\theta)$ in 
axisymmetry) and in the Regge-Wheeler gauge~\cite{Regge:1957td}:

%
\begin{align}
\label{eq:hpert}
&h_{\mu\nu}^{(n)}=
\left(\begin{array}{cccc}
-fH_0^{n\ell}P_\ell & 0 & 0 & h_0^{n\ell}P'_\ell \\
 0 & f^{-1}H_2^{n\ell}P_{\ell} & 0 &0\\
 0 &0 & r^2 K^{n\ell}P_{\ell}& 0\\ 
h_0^{n\ell }P'_\ell&0 &0 & r^2\sin^2{\theta} K^{n\ell}P_\ell
\end{array}\right)\,,
\end{align}
%
with ${P_\ell}'=\frac{dP_\ell(\cos\theta)}{d\cos\theta}$ and ${f(r)=1-2{\cal M}/r}$. 
The parameter $\ell$ is related to the multipolar series in the Legendre polynomials
(a sum over $\ell$ is implicit), whereas $n$ denotes the order of the 
perturbative scheme. It is convenient to separate the perturbations in two sets, 
according to how they transform under parity. The odd (or axial) sector 
contains only the function $h_0^{n\ell }(r)$, whereas the even (or polar) 
sector contains the functions $H_0^{n\ell}(r)$, $H_2^{n\ell}(r)$, and 
$K^{n\ell}(r)$. Owing to the harmonic decomposition, all these functions depend only 
on the radial coordinate $r$.

\subsection{Separation of variables}
By inserting metric~\eqref{eq:pertmetric} into the vacuum equations, 
$R_{\mu\nu}=0$, we obtain a set of ordinary differential equations for the perturbation 
functions $h^{(n)}_{\mu\nu}(r)$. We solve them analytically by using an 
extension of the perturbative scheme of Ref.~\cite{Maselli:2015tta}.

The odd parity sector is entirely characterized by the functions $h_0^{n\ell }$. At each 
given order $n\geq 1$ the differential equations can be obtained from $R_{t\varphi}=0$ by 
using the orthogonality of the axisymmetric vector spherical harmonics,
\begin{align}
\label{eq:odd}
\int_0^\pi d\theta\, \sin\theta {P_\ell}'(\cos\theta)R_{t\varphi} =0\,,
\end{align}
with $\ell=1,\,2,\,3,\, ...$ Due to the symmetry of the background and of the Einstein's 
field equations, this procedure gives a set of purely radial, ordinary differential 
equations for $h^{n \ell}_{0}$. These equations are inhomogeneous (for $n\geq 2$) with 
source terms given by the lower-order functions.  
 

The even parity sector is characterized by the functions $H_0^{n\ell}$, $H_2^{n\ell}$ and 
$K^{n\ell}$. Similarly to the odd parity case, at each given order $n\geq 1$ the 
differential equations can be obtained from $E^i\equiv(R_{tt},R_{rr},R_{\theta\theta})=0$ 
and by using the orthogonality of the Legendre polynomials $P_\ell$,
\begin{align}
\label{eq:even}
\int_0^\pi \sin\theta P_\ell(\cos\theta)E^i\,d\theta =0\,,
\end{align}
with $\ell=0,\,1,\,2,\,3,\, ...$  
This 
procedure leads to a set of 
purely radial, coupled, inhomogeneous differential equations for $H_0^{n\ell}$, 
$H_2^{n\ell}$ and $K^{n\ell}$, where again the source terms are given by the 
lower order functions.

\subsection{Solutions and multipole moments}
\begin{table}
\centering
\begin{tabular}{l|l}
\bf{Symbol} & 	\bf{Definition}\\
\hline
\hline
$\mathcal{M}$& Physical mass \\
$\mathcal{J}$& Physical angular momentum    \\
$\mathcal{M}_\ell$& Physical mass $\ell$-pole    \\ 
$\mathcal{S}_\ell$ & Physical current $\ell$-pole   \\ 
$M_\ell$ &  Leading-order mass $\ell$-pole   \\
$S_\ell$ & Leading-order current $\ell$-pole   \\ 
$M^{(n)}_\ell$ &  $n$th-order correction to the mass $\ell$-pole   \\
$S^{(n)}_\ell$ & $n$th-order correction to the current $\ell$-pole\\
\end{tabular}
\caption{Nomenclature used to describe the quantities of the quasi-Schwarzschild 
metric. Calligraphic symbols are used to represent physical quantities, while Latin 
symbols represent expansion coefficients of the same quantity. The subscript $\ell$ 
denotes the multipole order and the superscript $(n)$ represents the order of the 
correction.}
\label{nomenclature}
\end{table}

The set of ordinary differential equations provided 
in~\eqref{eq:odd} and~\eqref{eq:even} can be solved analytically for the metric 
functions to any given order. 
As we shall discuss, the structure of the solution is very similar order by 
order, so we expect that the following results will hold to any order in the 
perturbative scheme. 
This is due to the fact that each solution is a polynomial in ${\cal M}/r$ and 
also contains terms such as $\log\left(1-{2{\cal M}/r}\right)$ and 
powers thereof [see Eq.~\eqref{functiondecomposition} below]. When appearing in the source 
of the differential equation for the higher-order corrections, these terms 
give rise to the same polynomial and logarithmic terms. The procedure continues 
iteratively, although the higher-order solutions are cumbersome.
Table~\ref{nomenclature} summarizes the notation used in this section. 
In the following we briefly outline how to obtain the solution and extract the multipole 
moments.

\subsubsection{$\mathcal{O}(\epsilon)$ terms and $\ell\geq 2$}
At linear order, from Eq.~\eqref{eq:odd} we obtain a set of differential equations for 
the functions $h_0^{1\ell}$ and, from Eq.~\eqref{eq:even}, a set of differential equations 
for the functions $H_0^{1\ell}$, $H_2^{1\ell}$, $K^{1\ell}$. For each value of $\ell\geq 
2$, these systems can be reduced to a single equation for $H_0^{1\ell}$ and to a single 
equation for $h_0^{1\ell}$:
\begin{align}
&{\cal D}_1 H_0^{1\ell} =0\,,\\
&{\cal D}_2 h_0^{1\ell} =0\,,
\end{align}
where ${\cal D}_1$ and ${\cal D}_2$ are two second-order differential operators given by 
\begin{align}
&{\cal D}_1\equiv\frac{d^2}{dr_*^2}+\frac{2f}{r}\frac{d}{dr_*} 
-\left(f\frac{l(l+1)}{r^2}+\frac{4{ \cal M}^2}{r^4}\right)\,,\\
&{\cal D}_2\equiv\frac{d^2}{dr^2}+\frac{4 {\cal M}-l (l+1) r}{r^2 (r-2 {\cal M})}\,,
\end{align}
and $r_*$ is the tortoise coordinate defined by $dr/dr_*=f$. 
The above equations can be solved analytically, each function being defined by two 
arbitrary constants. The first constant can be fixed by imposing asymptotic flatness, 
whereas the second constant is related to the corresponding multipole moment of order 
$\ell$. The large-distance behaviour of the solutions reads
\begin{align}
\label{eq:h20}
&H_0^{1\ell}\to-\frac{2 M_\ell^{(1)}}{r^{\ell+1}}+O\left(r^{-(\ell+2)}\right)\,,\\
&h_0^{1\ell}\to-\frac{2}{\ell} \frac{S_\ell^{(1)}}{r^{\ell}}+O\left(r^{-(\ell+1)}\right)\,
\end{align}
where $M_\ell^{(1)} (S_\ell^{(1)})$ can be identified with the mass 
(current) $\ell$-pole moment to linear order in the perturbation scheme.

\subsubsection{$\mathcal{O}(\epsilon^n)$ terms and $\ell \geq 2$}
To $\mathcal{O}(\epsilon^n)$ the functions that appear in the metric are $H_0^{n\ell}$, 
$H_2^{n\ell}$, $K^{n\ell}$ and $h_0^{n\ell}$. As we discussed above, the equations for 
these solutions are sourced by the multipoles in the lower order terms. The differential 
equations can be written as,
\begin{align} \label{eqINH2}
&{\cal D}_1 H_0^{n\ell}={\cal T}_1^{n\ell}\,,
&{\cal D}_2 h_0^{n\ell}={\cal T}_2^{n\ell}\,,
\end{align} 
where ${\cal T}_1^{n\ell}$ and  ${\cal T}_2^{n\ell}$ are the source terms generated by 
the lower order moments. 

To any order the new solution is defined by a free constant associated with the 
homogeneous solution of Eq.~\eqref{eqINH2}, which is related to the corresponding 
multipole moments.
In the asymptotic limit, the inhomogeneous solutions read
\begin{align}
\label{eq:H0nl}
&H_0^{n\ell}\to-\frac{2 M_\ell^{(n)}}{r^{\ell+1}}+O\left(r^{-\ell-2}\right)\,,\\
\label{eq:h0nl}
&h_0^{n\ell}\to-\frac{2}{\ell} \frac{S_\ell^{(n)}}{r^{\ell}}+O\left(r^{-\ell-1}\right)\,,
\end{align}
where $M_\ell^{(n)} (S_\ell^{(n)})$ is an $n$th-order correction to the object's 
mass~(current) $\ell$-pole moment. 
These quantities are in general free constants that are proportional to the multipole 
moments which source them.

The analytical expressions of the metric functions are too cumbersome to 
present them explicitly here but they are provided in an online 
repository~\cite{webpage}. For any $n$, 
their schematic form reads
\begin{equation}
\label{functiondecomposition}
x_\ell^{(n)}=\sum_{i=0}^ {n} a^{(i)}(r)\log(1-2{\cal M}/r)^{i}
\end{equation}
where $x_\ell^{(n)}$=($H_0^{n\ell}$, $H_2^{n\ell}$, $K^{n\ell}$, $h_0^{n\ell}$) 
collectively represents all variables,
and $a^{(i)}(r)$ are generic polynomials in ${\cal M}/r$ which are regular at $r=2{\cal 
M}$ and depend on the multipole moments of the object.

\subsubsection{$\ell=0,1$}
For the axial sector of the perturbations, the structure of the solutions described above 
is still valid even when $\ell=1$. However, in the polar sector the case of 
$\ell=0,1$ is different. For ${\ell=0}$ we can set $K^{n0}=0$ for any $n$ without loss 
of generality, since the metric 
is invariant under a transformation ${r\to A(r) \,r}$. On the other hand, for $\ell=1$ the 
system of ordinary differential equations is underdetermined. This is 
a consequence of the fact that the $\ell=1$ polar perturbations include a 
displacement of the center of mass, which can be compensated by an appropriate choice of 
coordinates. To fix this remaining freedom we choose the constant 
$K^{n1}=0$ for any $n$.
With this choice one can solve the system of Eqs.~\eqref{eq:odd}~and~\eqref{eq:even}. The 
equations for the $\ell=0$ and $\ell=1$ components of the polar sector can be written 
schematically as
\begin{align}
&\frac{d^2 H_0^{n0}}{dr^2}+\frac{2 H_0^{n0}}{r-2 {\cal M}}={\cal T}_1^{n0}\,,\\
&\frac{dH_0^{n1}}{dr}+\frac{2 H_0^{n1}}{r-2 {\cal M}}={\cal T}_1^{n1}\,,
\end{align}
where the source terms are zero when $n=1$. The functions $H_2^{n0}$ and $H_2^{n1}$ 
can be written in terms of $H_0^{n0}$ and $H_0^{n1}$. 
\subsubsection{Truncating the multipolar orders}

Although our approach does not require to restrict to any given multipolar 
order, the multipolar expansion needs to be truncated in 
order to obtain a finite set of equations. The order of the truncation is in 
general arbitrary and depends on the particular solution. For example, in 
Sec.~\ref{sec:slowrot} 
below we shall restrict to \emph{spin-induced} multipole moments and truncate the 
multipolar order by imposing that only $\ell=1$ perturbations are nonvanishing 
at ${\cal O}(\epsilon)$, whereas $\ell>1$ perturbations are sourced at higher 
order.

More in general, the couplings between multipoles follow the standard 
addition rules for angular momenta in quantum mechanics, so that if one mode 
with $\ell_1$ is present at ${\cal O}(\epsilon)$ and another mode with $\ell_2>\ell_1$ is present at a ${\cal O}(\epsilon^n)$, they will source a multipole moment to ${\cal O}(\epsilon^{n+1})$ with $\ell$ such that 
$\ell_2-\ell_1\leq \ell\leq\ell_2+\ell_1$, provided some order is not forbidden by the 
selection rules described in Sec.~\ref{sec:properties}. (For a related discussion, 
see Ref.~\cite{Pani:2013pma}.)
For example, if only $\ell=1$ is present to ${\cal O}(\epsilon)$, it will source 
$\ell=0,1,2$ to ${\cal O}(\epsilon^2)$, and $\ell=0,1,2,3$ to ${\cal 
O}(\epsilon^3)$, and so on (this particular case is simply the Hartle-Thorne 
small-spin expansion~\cite{Hartle:1967he,1968ApJ...153..807H}, cf.\ 
Sec.~\ref{sec:slowrot}). 
On the other hand, if only $\ell=1,2$ are present to ${\cal O}(\epsilon)$, they will 
source $\ell=0,1,2,3,4$ to ${\cal O}(\epsilon^2)$, and $\ell=0,1,2,3,4,5,6$ to ${\cal 
O}(\epsilon^3)$, and so on.
In general, if the solution contains moments up to 
multipolar index $\ell=L$ to the leading order, then moments up to index 
$\ell=n\times L$ will be induced to ${\cal O}(\epsilon^n)$.

Thus, building a consistent solution requires to include at each given order $n$ in 
$\epsilon\ll1$ a sufficiently high number of terms in the multipolar 
expansion, depending on initial assumptions for the system under consideration. As we 
shall discuss, for some solutions the number of terms in the ansatz can be 
significantly reduced by making use of some parity rules.

\section{Properties of the quasi-Schwarzschild soft ECO metric}\label{sec:properties}


\begin{table*}[ht!]
\centering
\begin{tabularx}{0.85\textwidth}{Y||c|c|c|c|c}
 & {\centering $\epsilon^{1}$} & $\epsilon^{2}$ & $\epsilon^{3}$ & $\epsilon^{4}$ & $\epsilon^{5}$ \\
\hline
\hline
$\ell=0$ &\multicolumn{1}{@{}X@{}}{\tikzmark[31]{\raisebox{-1.5ex}{$0$}\raisebox{1.5ex}{\hspace{5ex}$M_0^{(1)}$}}}  &\multicolumn{1}{|@{}X@{}}{\tikzmark[32]{\raisebox{-1.5ex}{$0$}\raisebox{1.5ex}{\hspace{5ex}$M_0^{(2)}$}}} &\multicolumn{1}{|@{}X@{}}{\tikzmark[33]{\raisebox{-1.5ex}{$0$}\raisebox{1.5ex}{\hspace{5ex}$M_0^{(3)}$}}} &\multicolumn{1}{|@{}X@{}}{\tikzmark[34]{\raisebox{-1.5ex}{$0$}\raisebox{1.5ex}{\hspace{5ex}$M_0^{(4)}$}}}&\multicolumn{1}{|@{}X@{}}{\tikzmark[35]{\raisebox{-1.5ex}{$0$}\raisebox{1.5ex}{\hspace{5ex}$M_0^{(5)}$}}} \\
\hline
$\ell=1$ &\multicolumn{1}{@{}X@{}}{\tikzmark[1]{\raisebox{-1.5ex}{$S_1^{(1)}$}\raisebox{1.5ex}{\hspace{5ex}$0$}}}  &\multicolumn{1}{|@{}X@{}}{\tikzmark[2]{\raisebox{-1.5ex}{$S_1^{(2)}$}\raisebox{1.5ex}{\hspace{5ex}$0$}}} &\multicolumn{1}{|@{}X@{}}{\tikzmark[3]{\raisebox{-1.5ex}{$S_1^{(3)}$}\raisebox{1.5ex}{\hspace{5ex}$0$}}} &\multicolumn{1}{|@{}X@{}}{\tikzmark[4]{\raisebox{-1.5ex}{$S_1^{(4)}$}\raisebox{1.5ex}{\hspace{5ex}$0$}}}&\multicolumn{1}{|@{}X@{}}{\tikzmark[5]{\raisebox{-1.5ex}{$S_1^{(5)}$}\raisebox{1.5ex}{\hspace{5ex}$0$}}} \\
\hline
$\ell=2$ & \multicolumn{1}{@{}X@{}}{\tikzmark[6]{\raisebox{-1.5ex}{$S_2^{(1)}$}\raisebox{1.5ex}{\hspace{2ex}$M_2^{(1)}$}}} &\multicolumn{1}{|@{}X@{}}{\tikzmark[7]{\raisebox{-1.5ex}{$S_2^{(2)}$}\raisebox{1.5ex}{\hspace{2ex}$M_2^{(2)}$}}}  &\multicolumn{1}{|@{}X@{}}{\tikzmark[8]{\raisebox{-1.5ex}{$S_2^{(3)}$}\raisebox{1.5ex}{\hspace{2ex}$M_2^{(3)}$}}} &\multicolumn{1}{|@{}X@{}}{\tikzmark[9]{\raisebox{-1.5ex}{$S_2^{(4)}$}\raisebox{1.5ex}{\hspace{2ex}$M_2^{(4)}$}}}  &\multicolumn{1}{|@{}X@{}}{\tikzmark[10]{\raisebox{-1.5ex}{$S_2^{(5)}$}\raisebox{1.5ex}{\hspace{2ex}$M_2^{(5)}$}}} \\
\hline 
 $\ell=3$&  \multicolumn{1}{@{}X@{}}{\tikzmark[11]{\raisebox{-1.5ex}{$S_3^{(1)}$}\raisebox{1.5ex}{\hspace{2ex}$M_3^{(1)}$}}}  &  \multicolumn{1}{|@{}X@{}}{\tikzmark[12]{\raisebox{-1.5ex}{$S_3^{(2)}$}\raisebox{1.5ex}{\hspace{2ex}$M_3^{(2)}$}}}  &  \multicolumn{1}{|@{}X@{}}{\tikzmark[13]{\raisebox{-1.5ex}{$S_3^{(3)}$}\raisebox{1.5ex}{\hspace{2ex}$M_3^{(3)}$}}}  &   \multicolumn{1}{|@{}X@{}}{\tikzmark[14]{\raisebox{-1.5ex}{$S_3^{(4)}$}\raisebox{1.5ex}{\hspace{2ex}$M_3^{(4)}$}}} &  \multicolumn{1}{|@{}X@{}}{\tikzmark[15]{\raisebox{-1.5ex}{$S_3^{(5)}$}\raisebox{1.5ex}{\hspace{2ex}$M_3^{(5)}$}}} \\
\hline 
 $\ell=4$& \multicolumn{1}{@{}X@{}}{\tikzmark[16]{\raisebox{-1.5ex}{$S_4^{(1)}$}\raisebox{1.5ex}{\hspace{2ex}$M_4^{(1)}$}}} &\multicolumn{1}{|@{}X@{}}{\tikzmark[17]{\raisebox{-1.5ex}{$S_4^{(2)}$}\raisebox{1.5ex}{\hspace{2ex}$M_4^{(2)}$}}} & \multicolumn{1}{|@{}X@{}}{\tikzmark[18]{\raisebox{-1.5ex}{$S_4^{(3)}$}\raisebox{1.5ex}{\hspace{2ex}$M_4^{(3)}$}}}& \multicolumn{1}{|@{}X@{}}{\tikzmark[19]{\raisebox{-1.5ex}{$S_4^{(4)}$}\raisebox{1.5ex}{\hspace{2ex}$M_4^{(4)}$}}} & \multicolumn{1}{|@{}X@{}}{\tikzmark[20]{\raisebox{-1.5ex}{$S_4^{(5)}$}\raisebox{1.5ex}{\hspace{2ex}$M_4^{(5)}$}}} \\
\hline 
 $\ell=5$& \multicolumn{1}{@{}X@{}}{\tikzmark[21]{\raisebox{-1.5ex}{$S_5^{(1)}$}\raisebox{1.5ex}{\hspace{2ex}$M_5^{(1)}$}}} & \multicolumn{1}{|@{}X@{}}{\tikzmark[22]{\raisebox{-1.5ex}{$S_5^{(2)}$}\raisebox{1.5ex}{\hspace{2ex}$M_5^{(2)}$}}} &\multicolumn{1}{|@{}X@{}}{\tikzmark[23]{\raisebox{-1.5ex}{$S_5^{(3)}$}\raisebox{1.5ex}{\hspace{2ex}$M_5^{(3)}$}}} & \multicolumn{1}{|@{}X@{}}{\tikzmark[24]{\raisebox{-1.5ex}{$S_5^{(4)}$}\raisebox{1.5ex}{\hspace{2ex}$M_5^{(4)}$}}} & \multicolumn{1}{|@{}X@{}}{\tikzmark[25]{\raisebox{-1.5ex}{$S_5^{(5)}$}\raisebox{1.5ex}{\hspace{2ex}$M_5^{(5)}$}}}
\end{tabularx}

\caption{Representation of the multipolar structure of a soft ECO (described by the 
metric~\eqref{eq:pertmetric}) up to ${\cal O}(\epsilon^5)$ and $\ell\leq5$. Each 
column corresponds to the different multipole moments at a given perturbation 
order, while each line contains the different higher-order corrections to a 
given $\ell$-pole. Each cell entry $(n,\ell)$ is divided into two: the 
upper-right (lower-left) entry corresponds to the $n$-th order coefficient of 
the mass (current) $\ell$-pole, $M_\ell^{(n)}$ ($S_\ell^{(n)}$). Since our 
ansatz for metric~\eqref{eq:pertmetric} does not assume equatorial symmetry, all 
the entries in this table are present in the solution, with the exception of 
$M_1^{(n)}$, which can always be set to zero without loss of generality. The blue 
(red) cells correspond to the entries present in the Hartle-Thorne 
(Manko-Novikov) solution, whereas cells which are half blue and half red represent 
entries present in both of these particular cases.}
\begin{tikzpicture}[overlay,remember picture]
\draw[](1.north west)--(1.south east);
\draw[](2.north west)--(2.south east);
\draw[](3.north west)--(3.south east);
\draw[](4.north west)--(4.south east);
\draw[](5.north west)--(5.south east);
\draw[](6.north west)--(6.south east);
\draw[](7.north west)--(7.south east);
\draw[](8.north west)--(8.south east);
\draw[](9.north west)--(9.south east);
\draw[](10.north west)--(10.south east);
\draw[](11.north west)--(11.south east);
\draw[](12.north west)--(12.south east);
\draw[](13.north west)--(13.south east);
\draw[](14.north west)--(14.south east);
\draw[](15.north west)--(15.south east);
\draw[](16.north west)--(16.south east);
\draw[](17.north west)--(17.south east);
\draw[](18.north west)--(18.south east);
\draw[](19.north west)--(19.south east);
\draw[](20.north west)--(20.south east);
\draw[](21.north west)--(21.south east);
\draw[](22.north west)--(22.south east);
\draw[](23.north west)--(23.south east);
\draw[](24.north west)--(24.south east);
\draw[](25.north west)--(25.south east);  
\draw[](31.north west)--(31.south east);
\draw[](32.north west)--(32.south east);
\draw[](33.north west)--(33.south east);
\draw[](34.north west)--(34.south east);
\draw[](35.north west)--(35.south east);
\path[fill=blue,opacity=0.30](1.north west)--(1.south west) -- (1.center) -- cycle;
\path[fill=red,opacity=0.15](1.south east)--(1.south west) -- (1.center) -- cycle;
\path[fill=blue,opacity=0.30](3.north west)--(3.south west) -- (3.south east) -- cycle;
\path[fill=blue,opacity=0.30](5.north west)--(5.south west) -- (5.south east) -- cycle;
\path[fill=red,opacity=0.15](7.north east)--(7.south east) -- (7.center) -- cycle;
\path[fill=blue,opacity=0.30](7.north east)--(7.center) -- (7.north west) -- cycle;
\path[fill=blue,opacity=0.30](9.north east)--(9.south east) -- (9.north west) -- cycle;
\path[fill=blue,opacity=0.30](13.north west)--(13.south west) -- (13.center) -- cycle;
\path[fill=red,opacity=0.15](13.center)--(13.south west) -- (13.south east) -- cycle;
\path[fill=blue,opacity=0.30](15.north west)--(15.south west) -- (15.south east) -- cycle;
\path[fill=red,opacity=0.15](17.north east)--(17.south east) -- (17.north west) -- cycle;
\path[fill=red,opacity=0.15](19.north east)--(19.south east) -- (19.center) -- cycle;
\path[fill=blue,opacity=0.30](19.north east)--(19.center) -- (19.north west) -- cycle;
\path[fill=red,opacity=0.15](23.north west)--(23.south west) -- (23.south east) -- cycle;
\path[fill=blue,opacity=0.30](25.north west)--(25.south west) -- (25.center) -- cycle;
\path[fill=red,opacity=0.15](25.center)--(25.south west) -- (25.south east) -- cycle;
\path[fill=blue,opacity=0.30](32.north east)--(32.south east) -- (32.north west) -- cycle;
\path[fill=blue,opacity=0.30](34.north east)--(34.south east) -- (34.north west) -- cycle;
\end{tikzpicture}
\label{metric_graphical}
\end{table*}


A useful graphical representation of the multipolar structure of the general solution 
is given in Table~\ref{metric_graphical}. Each row represents the multipole moments of 
order 
$\ell$, while each column represents the ${\cal O}(\epsilon^n)$ correction to a 
given multipole. 
We can identify each entry of Table~\ref{metric_graphical} with a set 
$(n,\ell)$ corresponding to its parameters. Thus, each cell $(n,\ell)$ 
represents a $n$th-order correction to a multipole moment of order $\ell$. 
Furthermore, each 
entry is divided into two subcells. The upper right part 
represents the mass multipole moment contribution, whereas the bottom left part 
represents the current multipole moment contribution. The solution given by 
ansatz~\eqref{eq:pertmetric} contains all multipolar 
contributions represented in Table~\ref{metric_graphical}. 
Each line of 
Table~\ref{metric_graphical} can be reduced to two physical parameters, ${\cal 
M}_\ell$ and ${\cal S}_\ell$, describing the physical mass and current 
multipoles of that order\footnote{The row corresponding to $\ell=1$ is 
characterized by one parameter only, since the mass dipole can be set zero 
without loss of generality.}.

In addition to the sum rules for the multipolar index $\ell$ at each 
perturbative order, our framework also enjoys two sets of symmetries, which are related 
to the properties of each term under a parity transformation and under a reflection on 
the equatorial plane, respectively:
\begin{itemize}
 \item \emph{Parity rule}: Since mass (current) multipole moments are related with 
even (odd) functions in the metric decomposition 
[cf.~Eqs~\eqref{eq:H0nl}--\eqref{eq:h0nl}], terms associated to mass multipole moments 
are parity-even, 
whereas those associated to current multipole moments are parity-odd. The coupling between 
two parity-even (-odd) terms gives rise to a parity even term, whereas the coupling 
between a parity-even and a parity-odd term gives rise to a parity-odd term.
 \item \emph{$\mathbb{Z}_2$ rule}: Terms associated to ${\cal M}_\ell$ (${\cal S}_\ell$) 
are symmetric under a reflection on the equatorial plane when $\ell$ is even (odd), 
whereas they change sign when $\ell$ is odd (even). Thus, as previously mentioned, 
equatorial symmetry of the full 
metric implies ${\cal M}_\ell=0$ (${\cal S}_\ell=0$)  when $\ell$ is odd 
(even). In the general case, these restrictions do not hold and the
solution enjoys a simple ``$\mathbb{Z}_2$ rule'': the coupling between two terms that 
are both even or odd under this transformation gives rise to a $\mathbb{Z}_2$-even term, 
whereas the coupling between an even and an odd term gives rise to a $\mathbb{Z}_2$-odd 
term. In other words, the coupling between two (non)equatorially-symmetric moments gives 
rise to equatorially-symmetric moments, whereas the coupling between an 
equatorially-symmetric moment and a nonequatorial one gives rise to 
nonequatorially-symmetric moments.
\end{itemize}

The above rules, together with the addition rules of angular momenta, strongly 
constrain the types of multipole moments that can be induced for each perturbative 
solution. 
For example, to order ${\cal O}(\epsilon^2)$ a ${\cal M}_4$ moment cannot be induced 
by the coupling between ${\cal M}_2$ and ${\cal S}_2$, even if the angular-momentum sum 
rules would allow for it. Indeed, both ${\cal M}_2$ and ${\cal M}_4$ are $\mathbb{Z}_2$ 
even while ${\cal S}_2$ is $\mathbb{Z}_2$ odd. Likewise, a ${\cal S}_3$ moment (which 
is parity odd) cannot be induced by ${\cal M}_2$ (which is parity even) to order ${\cal 
O}(\epsilon^2)$, even if both the angular-momentum sum and the $\mathbb{Z}_2$ rules would 
allow for it.
Explicit solutions in which all these rules are at play are discussed in 
Sec.~\ref{sec:particular} below.

As we shall discuss in the next section, our solution can be reduced to the
Hartle-Thorne small-spin expansion if we consider all multipole moments to be spin 
induced and take 
the spin to be ${\cal O}(\epsilon)$. In this particular case the only nonzero 
entries in Table~\ref{metric_graphical} are those identified in blue.
Due to the aforementioned $\mathbb{Z}_2$ rule, these terms are all equatorially 
symmetric since are sourced by powers of the spin ${\cal J}\equiv {\cal S}_1$, which is 
$\mathbb{Z}_2$ even.

It is also informative to compare our quasi-Schwarzschild solution to known exact vacuum 
solutions to Einstein's equation, for example with the Manko-Novikov metric~\cite{MN1991}
with arbitrarily large multipole moments~\cite{inprep}.
The Manko-Novikov solution has a multipolar 
structure that in Table~\ref{metric_graphical} can be identified by the red cells. As 
clear from the table, this solution is 
\emph{not} a particular case of the Hartle-Thorne approximation --~even in the 
limit of small multipole moments~-- but it is actually \emph{orthogonal} to it. This 
is consistent with the fact that, in the small-multipole moment limit, the 
Manko-Novikov solution contains quadratic spin terms in the multipole moments 
equal to or higher than $\ell=4$ (see, e.g., 
Ref.~\cite{Frutos-Alfaro:2016arb}), which is not the case for the Hartle-Thorne 
metric, since in that case ${\cal M}_\ell \sim \chi^\ell$ and ${\cal S}_\ell 
\sim 
\chi^\ell$ (at least) to the 
leading order~\cite{Hartle:1967he,1968ApJ...153..807H,Yagi:2014bxa,Maselli:2015tta} 
(see discussion below). Also in this case, the selection and $\mathbb{Z}_2$ 
rules discussed above enforce the solution to be equatorially symmetric.

Besides these particular cases, the solution resulting from 
ansatz~\eqref{eq:pertmetric} is generically not equatorially symmetric, as can be 
seen by the presence of even (odd) current (mass) multipole moments in 
Table~\ref{metric_graphical}. Some specific nonequatorially symmetric solutions are 
discussed in Sec.~\ref{sec:PC_noneq}.

\section{Particular solutions} \label{sec:particular}
%

We have explicitly computed six different solutions, which we divide in two sets. The 
first set (Sec.~\ref{sec:PC_eq}) are three equatorially-symmetric solutions: (i) a 
small-multipole solution up to ${\cal O}(\epsilon^5)$ assuming only nonvanishing ${\cal 
J}$ at linear order (cf.~Sec.~\ref{sec:slowrot}); (ii)~a solution up to ${\cal 
O}(\epsilon^3)$ built by assuming that only ${\cal M}_2$ is present at linear order (cf. 
Sec.~\ref{sec:sol_M2}); and (iii)~a more general solution up to ${\cal 
O}(\epsilon^2)$ which assumes that both ${\cal J}$ and 
${\cal M}_2$ are present at linear order (cf. Sec.~\ref{sec:sol_JM2}). The second set 
(Sec.~\ref{sec:PC_noneq}) comprises three nonequatorially symmetric solutions: (iv)~a 
solution up to ${\cal O}(\epsilon^3)$ built with the assumption that at linear order only 
the ${\cal S}_2$ multipole moment is present (cf. Sec~\ref{sec:sol_S2a}); (v) a solution 
up to ${\cal O}(\epsilon^2)$ assuming both ${\cal J}$ and ${\cal S}_2$ at linear order 
(cf.~Sec.~\ref{sec:sol_JM2}); (vi) a more general solution up to ${\cal 
O}(\epsilon^2)$ constructed with the combination of 
 ${\cal J}$, ${\cal M}_2$ and ${\cal S}_2$ at linear order (cf. Sec.~\ref{sec:slowrot4}). 
The explicit form of these metrics is provided in an online notebook~\cite{webpage}, 
whereas some illustrative embedding diagrams are given in Fig.~\ref{fig:embedding}.

Since in this perturbative scheme each solution is completely characterized by 
its leading-order corrections only, we shall denote it by the multipole 
moments that exist at the linear level. Schematically, we shall denote as $\left[{\cal 
M}_i {\cal S}_j {\cal M}_k {\cal S}_l \ldots \right]$ a solution which has ${\cal 
M}_i$, ${\cal M}_k$, ${\cal S}_j$, ${\cal S}_l$, etc nonvanishing multipole 
moments (besides the mass) at the linear order.
With this nomenclature, the six solutions described above are denoted as 
$\left[{\cal J}\right]$, $\left[{\cal M}_2\right]$, $\left[{\cal J M}_2\right]$, 
$\left[{\cal S}_2\right]$, $\left[{\cal JS}_2\right]$ and  $\left[{\cal J M}_2{\cal 
S}_2\right]$, respectively. We stress that the quantities within brackets are the 
multipole moments entering at the leading order, but the full solution will contain more 
moments that are induced by the fundamental ones.

We also remark that, beyond ${\cal O}(\epsilon)$, a linear combination of two 
solutions is not a solution of the 
field equations, e.g. 
we cannot simply combine the $\left[{\cal J}\right]$-solution and the $\left[{\cal 
M}_2\right]$-solution to obtain the $\left[{\cal J M}_2\right]$-solution. This is a 
consequence of the nonlinearity of Einstein's equations. For example, the $\left[{\cal J 
M}_2\right]$-solution contains terms proportional to ${\cal J}{\cal M}_2$ {at ${\cal 
O}(\epsilon^2)$}, which do not 
appear in a linear combination of the $\left[{\cal J}\right]$ and $\left[{\cal 
M}_2\right]$ solutions.

To simplify the notation, in the following we will use the \emph{dimensionless} multipole 
moments
\begin{equation}
 \bar{{\cal M}_\ell} \equiv\frac{{\cal M}_\ell}{{\cal M}^{\ell+1}}\,,\qquad \bar {\cal 
S}_\ell \equiv\frac{{\cal S}_\ell}{{\cal M}^{\ell+1}}\,. \label{dimensionless}
\end{equation}

\subsection{Equatorially symmetric case}\label{sec:PC_eq}
\subsubsection{$\left[{\cal J}\right]$-solution}\label{sec:slowrot}
Let us assume that the multipole moments are sourced only by the object's 
rotation. Thus, to linear order in the spin, corresponding to ${\cal 
O}(\epsilon)$, the multipolar structure of the object is modified only by means 
of the body's angular momentum, and all multipole moments with $\ell\geq2$ are 
spin induced. This allows us to truncate the 
multipole moments of $\mathcal{O}(\epsilon)$ at $\ell=1$, i.e. 
$H_0^{1\ell}=H_2^{1\ell}=K^{1\ell}=h_0^{1\ell}=0$ for any 
$\ell>1$. 
This solution corresponds to the Hartle-Thorne approximation for the external spacetime 
of slowly-rotating objects in general 
relativity~\cite{Hartle:1967he,1968ApJ...153..807H,Yagi:2014bxa}.

The properties of this solution are well-known but for the sake of 
clarity we will summarize them here. Due to the $\mathbb{Z}_2$ rules presented in 
Sec.~\ref{sec:properties}, this 
solution is equatorially symmetric. Furthermore, due to the addition rules of angular 
momenta, to second order in the spin, the object's angular momentum ($\ell=1$) 
will source a mass quadrupole ($\ell=2$) and a correction to the object's mass 
($\ell=0$). Similarly, to third order in the spin, it will source a current 
octupole ($\ell=3$) and a correction to the object's angular momentum 
($\ell=1$). The corrections to the lower-order moments (mass and angular 
momentum) can be reabsorbed by defining the physical quantities ${\cal M}=M+\epsilon^2 
M_{0}^{(2)}+{\cal O}(\epsilon^4)$ and ${\cal J}=\epsilon J+\epsilon^3 S_{1}^{(3)}+{\cal 
O}(\epsilon^5)$.
We can summarize the structure of the equations in the 
following form: to each even (odd) order $n$, the spin will source a 
mass (current) multipole moment of order $\ell=n$, and will also source 
corrections to the lower mass (current) multipole moments.
Schematically, the spin-induced multipole moments of this solution 
read, to the \emph{leading} order\footnote{Generically, the spin-induced moments $\bar 
{\cal M}_\ell$ and $\bar {\cal S}_\ell$ might also get corrections at order 
$\chi^{\ell+2n}$ ($n=1,2,3,..$), but these are subleading in the small-spin expansion. 
Henceforth we shall denote the possible subleading corrections with ellipsis.} in the 
spin expansion:
\begin{align}
 \bar {\cal M}_\ell &\propto \chi^\ell+... \,, \qquad {\rm 
even}~\ell\geq2\\
 \bar {\cal S}_\ell &\propto \chi^\ell+... \,, \qquad {\rm odd}~\ell\geq2
\end{align}
where the prefactors are model-dependent constants that depend only on the compactness, 
${\cal M}_\ell=0$ (${\cal S}_\ell=0$) for odd (even) $\ell$, and we have defined the 
dimensionless moments as in Eq.~\eqref{dimensionless}.
This multipolar structure corresponds to the blue cells in Table~\ref{metric_graphical}.
We note that the exterior solution of a slowly-spinning object in general relativity has 
been computed up to ${\cal O}(\chi^4)$ in Ref.~\cite{Yagi:2014bxa}; our $\left[{\cal 
J}\right]$ metric extends that result to ${\cal O}(\chi^5)$ and might therefore be useful 
to construct more precise models of slowly-spinning neutron stars.

\subsubsection{$\left[{\cal M}_2\right]$-solution}\label{sec:sol_M2}
One of the simplest extensions to the previous model consists in assuming that the 
multipole moments are induced not by the spin but by the mass quadrupole moment.  
Similarly to the role of the spin in 
the case above (Sec.~\ref{sec:slowrot}), the 
quadrupole moment ${\cal M}_2$ will appear in the source terms of the 
higher-order 
equations.
For example, at second order in the expansion, the coupling between two 
mass quadrupole terms will source 
an ${\cal M}_4$ multipole moment and subleading corrections to the quadrupole and the 
mass. In a similar way, at third order the coupling between the multipoles will source an 
${\cal M}_6$ multipole moment and corrections to all the existing lower order multipole 
moments. As discussed previously, the subleading corrections to the multipole moments 
existing to the leading order (mass monopole and mass quadrupole) can be
reabsorbed in the definition of the \emph{physical} multipole moments.
In contrast to the $\left[{\cal J}\right]$-solution, this case does not 
contain current multipole moments. At each order $n$ 
only mass multipole moments up to order $\ell=2n$ will be sourced.
Schematically, the \emph{quadrupole-induced} multipole moments of this 
solution read:
\begin{align}
 \bar {\cal M}_{\ell} &\propto \left(\bar {\cal M}_2\right)^{\ell/2}+...\,,\qquad {\rm 
even}~\ell\geq4 \\
{\cal S}_\ell &=0\,, \qquad\qquad\qquad\quad~ {\rm any}~\ell\,,
\end{align}
whereas ${\cal M}_\ell=0$ for odd $\ell$. Again, the ellipsis in the above equation refer 
to the subleading corrections entering at higher order in the $\bar {\cal M}_2\ll1$ 
expansion, and the prefactors are model-dependent 
constants that depend only on the compactness. 

\subsubsection{$\left[{\cal J M}_2\right]$-solution}\label{sec:sol_JM2}
Another interesting solution comes from the combination of the two cases above, 
namely when both the spin and the quadrupole moment act as source terms.
Obviously this solution must reduce to the $\left[{\cal 
J}\right]$-solution ($\left[{\cal M}_2\right]$-solution) when we take ${\cal M}_2=0$ 
(${\cal J}=0$). However, it is not a simple linear combination of 
$\left[{\cal J}\right]$ and $\left[{\cal M}_2\right]$, since it also contains 
mixed coupling terms at ${\cal O}(\epsilon^2)$, i.e. terms proportional to ${\cal J} 
{\cal M}_2$. For 
example, at second order in the 
expansion, this coupling will source a current 
octupole moment, ${\cal S}_3$, and a higher-order correction to the object's angular 
momentum, which can be reabsorbed as in the $[{\cal J}]$ solution. For each order $n$ 
this 
solution contains mass multipole moments for even $\ell$ and current multipole moments 
for 
odd $\ell$ up to $\ell=2n$.

The induced multipole moments of this solution can be written as a combination 
of terms sourced by the spin and by the quadrupole:
\begin{align}
\label{eq:inducedmassJM2}
 \bar{\cal M}_{\ell} &:= \sum_{p=0}^{\ell/2}\alpha_{p} \chi^{\ell-2p}\left(\bar {\cal 
M}_2\right)^{p}+...\,, \qquad {\rm even}~\ell\geq4\\
 \label{eq:inducedcurrentJM2}
  \bar{\cal S}_\ell &:= \sum_{p=0}^{(\ell-1)/2}\beta_{p} \chi^{\ell-2p}\left(\bar {\cal 
M}_2\right)^{p}+...\,, \quad {\rm odd}~\ell\geq3\,,
\end{align}
whereas ${\cal M}_\ell=0$ (${\cal S}_\ell=0$) for odd (even) $\ell$. The 
prefactors $\alpha_{p}$ and $\beta_{p}$ are 
model-dependent, dimensionless constants that depend only on the compactness, and the 
ellipsis refers to subleading terms.

The above formulas can be understood as follows. The angular-momentum addition 
rules imply that, to the leading order, ${\cal M}_\ell\sim \chi^q ({\cal M}_2)^p$, with 
$q+2p=\ell$ since $\chi$ is a $\ell=1$ moment and ${\cal M}_2$ is a $\ell=2$ moment. 
By replacing $q=\ell-2p$ for each order $p$, we obtain Eq.~\eqref{eq:inducedmassJM2}. 
A similar argument applies to the derivation of Eq.~\eqref{eq:inducedcurrentJM2}.

For example, in this solution the (dimensionless) induced $\bar {\cal M}_4$ up 
to ${\cal O}(\epsilon^4)$ reads
\begin{equation}
 \bar {\cal M}_4 = \alpha_2 (\bar{\cal M}_2)^2 +\alpha_1 \chi^2 \bar{\cal 
M}_2+\alpha_0 \chi^4\,, \label{M4_JM2}
\end{equation}
where the second term on the right-hand side is an example of the aforementioned coupling 
terms between two different multipole moments existing at the leading order.
%

\subsection{Nonequatorially symmetric case} \label{sec:PC_noneq}

\subsubsection{$\left[{\cal S}_2\right]$-solution}\label{sec:sol_S2a}
The simplest nonequatorially symmetric solution can be obtained by including only a 
current quadrupole moment at the leading order; i.e., higher-order 
multipoles will be sourced exclusively by ${\cal S}_2$. Due to the angular momentum and 
the $\mathbb{Z}_2$ selection rules described in Sec.~\ref{sec:properties}, the 
current quadrupole moment will source, to second order in the perturbation expansion, the 
(equatorial-symmetric) moments ${\cal M}_4$ and ${\cal M}_2$, and also a subleading 
correction to ${\cal M}$.
 Using the same selection rules we find that third-order couplings between these 
multipole moments will source the 
nonequatorially symmetric terms ${\cal S}_4$, ${\cal S}_6$ and a subleading 
correction to ${\cal S}_2$.
As in the above cases, the subleading corrections to the fundamental moments (mass 
monopole and current quadrupole) can be consistently reabsorbed in 
the definition of the physical multipole moments ${\cal M}$ and ${\cal S}_2$.
Although we computed the explicit solution up to third order in $\bar {\cal S}_2\ll1$, 
from the structure of the solution and the selection rules it is easy to find a peculiar 
general pattern, namely that the current multipole moments appear only at odd orders in 
the expansion while the mass multipole moments are sourced only at 
even orders in the expansion. Furthermore, it can also be shown that odd order terms are 
nonequatorially symmetric while even order terms are equatorially symmetric.  

Since this solution sources two new multipole moments at each perturbation 
order, 
a schematic form for current quadrupole-induced multipole moments is more involved than 
in the previous cases. Schematically, we obtain
\begin{align}
\bar{\cal M}_{\ell} &\propto \left(\bar{\cal S}_2\right)^{\lfloor\ell/2+1,2\rfloor}+... 
\qquad 
{\rm even}~\ell\geq2 \\
\bar{\cal S}_{\ell} &\propto \left(\bar{\cal 
S}_2\right)^{\lfloor\ell/2,2\rfloor+1}+... 
\qquad 
{\rm even}~\ell\geq4 
\end{align}
whereas ${\cal M}_{\ell}={\cal S}_{\ell}=0$ for odd values of $\ell$. In the above 
expressions $\lfloor x,2\rfloor:={\rm max}\lbrace 2n \in \mathbb{Z} \,| \,2n\leq 
x\rbrace$.
More explicitly, the above relations imply the multipole moments are sourced in 
pairs: 
$\bar{\cal M}_{2} \propto \left(\bar{\cal S}_2\right)^2$, 
$\bar{\cal M}_{4} \propto \left(\bar{\cal S}_2\right)^2$, 
$\bar{\cal M}_{6} \propto \left(\bar{\cal S}_2\right)^4$, 
$\bar{\cal M}_{8} \propto \left(\bar{\cal S}_2\right)^4$,
and likewise 
$\bar{\cal S}_{4} \propto \left(\bar{\cal S}_2\right)^3$,
$\bar{\cal S}_{6} \propto \left(\bar{\cal S}_2\right)^3$, 
$\bar{\cal S}_{8} \propto \left(\bar{\cal S}_2\right)^5$, 
$\bar{\cal S}_{10}\propto \left(\bar{\cal S}_2\right)^5$, 
and so on.
\subsubsection{$\left[{\cal JS}_2\right]$-solution}\label{sec:slowrot3}
In this case, the 
quadrupole moment ${\cal S}_2$ plays the same role of the mass quadrupole 
moment in the equatorially-symmetric case $\left[{\cal J 
M}_2\right]$ (Sec.~\ref{sec:sol_JM2}), and will appear in the source 
terms of the higher-order 
equations. For example, at second order in the perturbation expansion, the coupling 
between the (nonequatorial-symmetric) current quadrupole moment ${\cal S}_2$ and the 
(equatorial-symmetric) angular momentum ${\cal J}$ will source\footnote{In 
general also ${\cal M}_1$ would be sourced, however, choosing specific values 
for 
the integration constants of the homogeneous part of the equation we can set 
${\cal M}_1$=0, as expected since the mass dipole can be gauged away in general 
relativity.} a (nonequatorial-symmetric) mass octupole moment ${\cal M}_3$.

With this result in mind we can easily generalize the structure of the $\left[{\cal 
JS}_2\right]$-solution
in the following way. It describes a quasi-spherical symmetric body 
without equatorialsymmetry with just two independent parameters (besides the mass): 
its spin ${\cal 
J}$, and its current quadrupole ${\cal S}_2$. To each even (odd) order $n>2$, these two 
multipole moments will source a mass (current) multipole moment of 
order $\ell=n$, and will also source corrections to all lower mass (current) multipole 
moments which can be consistently reabsorbed.
Up to ${\cal O}(\epsilon^4)$, the first induced multipole moments of this 
solution can be written as a 
combination of terms sourced by spin and by the current quadrupole:
\begin{align}
 \bar {\cal M}_{3} &= a_3\chi\left(\bar{\cal S}_2\right)\,,\\
 \bar {\cal M}_{4} &= c_4\left(\bar{\cal S}_2\right)^{2}+b_{4} \chi^{4}\,,\\
 \bar  {\cal S}_3 &= d_3\chi^3  \,,\\
 \bar  {\cal S}_4 &= f_4 \left(\bar{\cal S}_2\right)^3 +g_4\chi^2 \left(\bar{\cal 
S}_2\right)\,,
\end{align}
where the prefactors $a_{\ell}$, $b_{\ell}$, $c_{\ell}$ and $d_{\ell}$, $f_\ell$ and 
$g_\ell$ are again model-dependent constants that depend only on the compactness.

\subsubsection{$\left[{\cal J M}_2{\cal S}_2\right]$-solution}\label{sec:slowrot4}

The most general solution which contains all multipole moments with $\ell\leq2$ 
at the 
leading order is the $\left[{\cal J M}_2{\cal 
S}_2\right]$-solution. This solution is characterized by three independent 
``hairs'' (in addition to the mass) --~the  angular momentum, the 
mass quadrupole moment, and the current quadrupole moment~-- the combination of which 
will source higher-order moments accordingly to the aforementioned selection rules. 
Similarly to the two previous cases, the presence of a nonvanishing 
current quadrupole moment breaks the $\mathbb{Z}_2$ symmetry of the solution.
Indeed, all previous cases discussed so far are included in this solution and can be 
obtained by setting some of the independent parameters to zero (e.g. the $\left[{\cal 
J M}_2{\cal S}_2\right]$-solution reduces to $\left[{\cal J M}_2\right]$ if we set ${\cal 
S}_2=0$). However, as mentioned before, a linear combination of the previous models is 
not 
sufficient to describe this solution nor will it be in general a solution of Einstein's 
field equations.

The structure 
of the solution is similar to the previous ones with the exception of new coupling terms 
that appear at higher order, as a consequence of the
nonlinearity of Einstein's equations. In addition to the couplings described in 
the former cases, this solution will contain a mixed coupling between ${\cal S}_2$ and 
${\cal M}_2$ which will source higher-order multipole moments. Due to the selection rules 
described in Sec.~\ref{sec:properties}, the coupling between the former 
(nonequatorial-symmetric) and the latter (equatorial-symmetric) will induce two 
nonequatorially symmetric terms with $\ell=2$ and $\ell=4$ at second order in the 
perturbative expansion. Thus, at second order the coupling ${\cal S}_2{\cal M}_2$ 
sources an ${\cal S}_4$ and a subleading (and re-absorbable) correction to ${\cal S}_2$.

Up to ${\cal O}(\epsilon^4)$, the first induced multipole moments of this 
solution can be written as a 
combination of terms sourced by spin, the mass quadrupole, and the current quadrupole:
\begin{align}
 \bar {\cal M}_{3} &=  a_{110}\chi\bar{\cal S}_2 +a_{111} \chi\bar{\cal S}_2\bar{\cal 
M}_2+a_{310}\chi^3\bar{\cal 
S}_2\nn\\
&+a_{130}\chi\left(\bar{\cal S}_2\right)^3+a_{112} \chi\bar{\cal S}_2\left(\bar{\cal 
M}_2\right)^2 
\,,\\
 \bar {\cal M}_{4} &= a_{020}\left(\bar{\cal 
S}_2\right)^{2}+a_{002}\left(\bar{\cal M}_2\right)^{2}+ a_{021}\left(\bar{\cal 
S}_2\right)^{2}\bar{\cal M}_2 \nn\\
& + a_{201} \chi^2 \bar{\cal M}_2+a_{400} \chi^{4}+a_{040}\left(\bar{\cal 
S}_2\right)^{4} +a_{004}\left(\bar{\cal M}_2\right)^{4}\nn\\
&+a_{220} \chi^2 \left(\bar{\cal 
S}_2\right)^{2}+ a_{202} \chi^2 \left(\bar{\cal M}_2\right)^{2}
+ a_{022}\left(\bar{\cal S}_2\right)^{2}\left(\bar{\cal M}_2\right)^2
\,,\\
 \bar {\cal S}_3   &=   a_{101} \chi \bar{\cal M}_2+ a_{300}\chi^3+ a_{102} \chi 
\left(\bar{\cal M}_2\right)^2+a_{120}\chi\left(\bar{\cal S}_2\right)^2
\nn\\
 &+ a_{301} \chi^3 \bar{\cal M}_2+ a_{103} \chi \left(\bar{\cal M}_2\right)^3
 +a_{121} \chi \left(\bar{\cal S}_2\right)^2\bar{\cal M}_2
\,,\\
 \bar {\cal S}_4  & = a_{011}\bar{\cal S}_2\bar{\cal M}_2+ a_{030}\left(\bar{\cal 
S}_2\right)^3 +a_{210}\chi^2 \bar{\cal S}_2\nn\\
 & +a_{012}\bar{\cal S}_2\left(\bar{\cal M}_2\right)^2+ a_{031}\left(\bar{\cal 
S}_2\right)^3\bar{\cal M}_2 + a_{211}\chi^2\bar{\cal S}_2\bar{\cal M}_2\,,
\end{align}
where the prefactors $a_{ijk}$ are again model-dependent constants that depend only on 
the compactness, and the subscripts are assigned such that the terms are of the form 
$a_{ijk}\chi^i\left(\bar{\cal S}_2\right)^j \left(\bar{\cal M}_2\right)^k
$.

\section{Hair conditioner for ECOs: soft hair in the BH limit} 
\label{sec:MM}
With the explicit form of the soft ECO solutions at hand we can now study the 
magnitude of the curvature on the surface of the ECO, and thus explore under what 
conditions the multipoles constitute soft hair.
To this end, we look at two curvature invariants, namely the 
Kretschmann scalar ${\cal K}\equiv R_{abcd}R^{abcd}$ and the Pontryagin scalar, 
${}^*RR\equiv\frac12 R_{abcd}\epsilon^{baef} {R^{cd}}_{ef}$. These are the leading-order 
nonvanishing curvature scalars, since our solutions satisfy $R_{\mu\nu}=0$.
The general expression of these invariants is cumbersome; however, they can be 
schematically decomposed into terms that are regular at $r=2{\cal 
M}$ and terms that are divergent. The terms containing a curvature singularity at 
$r=2{\cal M}$ must vanish identically in the BH case, yielding the unique Kerr solution.

However, the singularity is avoided if the object is just slightly less compact than a 
BH, i.e. if it has a radius $r_0>2{\cal 
M}$. This is the case (for instance) of perfect-fluid stars, whose deformations can also 
be described by our solution. An 
interesting question is therefore what happens for an object with $r_0=2{\cal 
M}(1+\delta)$ in the $\delta\to0$ limit.
Here, $r_0$ is the proper 
circumferential radius of a circle around the axis of 
symmetry~\cite{inprep}, so
\begin{equation}
 \delta = \frac{r_0}{2{\cal M}}-1 \label{def:delta1}
\end{equation}
is defined in terms of geometrical quantities.
Likewise, one could also introduce~\cite{Maselli:2018fay} the 
proper radial distance 
$\Delta=\int_{2{{\cal M}}}^{r_0}dr\,\sqrt{g_{rr}}\approx4{{\cal 
M}}\sqrt{\delta}+{\cal O}(\delta)$, which
is directly related to $\delta$ in a model-independent way in the $\epsilon\to0$ 
and $\delta\to0$ limits.

Multipole moments that would have to vanish in the Kerr limit are ECO hair. We will 
determine what their size can be, as a function of $\delta$, in order for the curvature 
of the solution not to be large (on the scale of $\cal M$). Under these conditions the 
ECO hair will remain soft.

\subsubsection{Multipole moments sourced by $\ell=1$ moments}\label{sec:curvatureJ}

As a starting point we focus on the case of spin-induced multipole moments. For this 
purpose we will use the $[{\cal J}]$- solution described in Sec.~\ref{sec:slowrot} and 
study its curvature invariants. Up to the linear order, the only multipole moments 
existing in the solution are the mass and the spin of the object and thus the curvature 
invariants are regular everywhere. At second order a spin-induced quadrupole moment is 
sourced. In the $r_0\to 2 {\cal M}$ limit, the second-order correction to the 
Kretschmann scalar at the surface reads
\begin{align}
&{\cal K}^{(2)}\sim-\frac{45\chi^2}{4 
{\cal M}^4}\left(1+A_2(\delta)\right)\log\left(1-\frac{2{\cal M}}{r_0}\right)P_2(\theta)
\,,
\end{align}
where we have explicitly factored out the spin dependence by defining 
\begin{equation}
\bar {\cal 
M}_2=A_2(\delta) \chi^2+{\cal O}(\chi^4),
\end{equation}
 with $A_2(\delta)$ being a function of 
$\delta$ only.
The logarithmic term above forces the curvature to blow 
up when $r_0\to 2{{\cal M}}$ unless $1+A_2(\delta)\to 0$ logarithmically, or 
faster. That is, $1+A_2(\delta)\to a_2/\log(\delta)$ or faster, where $a_2$ is a 
model-dependent constant at most of order unity.
In other words, requiring that the 
curvature does not blow up in the BH limit 
imposes
\begin{equation}
 \bar{\cal M}_2 \to \bar{\cal M}_2^{\rm 
BH}+a_2\frac{\chi^2}{\log\delta}\,, 
\label{dM2log}
\end{equation}
or faster as $\delta\to 0$, for a purely spin-induced solution. Here
\begin{equation}
 \bar{\cal M}_2^{\rm BH}=-\chi^2
\end{equation}
is the mass quadrupole moment of a Kerr BH [see Eq.~\eqref{nohair}].
Thus, as the ECO's surface approaches 
the horizon, the ECO's multipole moments must approach the BH multipole moments as 
$1/\log\delta$ or faster if the curvature is to remain of the same order as the curvature 
at the horizon of the corresponding BH (i.e., for a soft ECO).

By extending this analysis to higher-order terms we find that this behaviour occurs for 
all spin-induced multipole moments. As an example, to third order in the perturbation 
scheme the corrections to the Pontryagin scalar take the form\footnote{For the sake of 
clarity, in Eq.~\eqref{RRstar} we have assumed $\delta {\cal M}_2=0$. Nevertheless, the 
same conclusions can be reached if we assume that the quadrupole deviations vanish 
logarithmically, as requested by Eq.~\eqref{dM2log}.}\,,
\begin{align}
&{}^*RR^{(3)}\sim-\frac{315\chi^3}{4 
{\cal M}^{4}}  \left(1+C_3(\delta)\right)\log\left(1-\frac{2{\cal M}}{r_0}\right)P_3 
\,, 
\label{RRstar}
\end{align}
where similarly to the previous case we have defined 
\begin{equation}
{\bar{\cal 
S}_3=C_3(\delta)\chi^3+\mathcal{O}(\chi^5)}.
\end{equation}
Thus, the Pontryagin scalar blows up in the BH limit unless $1+C_3(\delta)\to 
c_3/\log(\delta)$ or faster, as $\delta\to 0$. In other words, the 
spin-induced current octupole moment of a ``soft'' ECO must vanish as 
\begin{equation}
\bar{\cal S}_3\to \bar{\cal S}_3^{\rm BH}+c_3\frac{\chi^3}{\log \delta}\,,
\end{equation}
or faster, as $\delta\to 0$. Again, $c_3$ is a new model-dependent constant of 
order unity or smaller, and
\begin{equation}
\bar{\cal S}_3^{\rm BH}=-\chi^3
\end{equation}
corresponds to the Kerr current octupole moment. 
	
These results lead to the conclusion that if an ECO is ``soft'' its spin-induced 
multipole moments must vanish logarithmically (or faster) as a function of $\delta\to0$, 
as anticipated by the first terms in 
Eqs.~\eqref{conjectureM} and \eqref{conjectureS}.

\subsubsection{Multipole moments sourced by $\ell>1$ moments}\label{sec:curvatureS2}
An interesting question is how do the induced multipoles vanish when they are 
sourced by multipole moments higher than the spin, e.g. by a leading-order quadrupole 
moment. Let us 
consider the nonequatorially symmetric $[{\cal S}_2]$-solution as a simple 
representative example. In this 
case the curvature invariants have 
pathologies at $r=2{\cal M}$ already at the linear order in the perturbative 
expansion,
\begin{align}
&{}^*RR^{(1)}\sim\frac{90 {\cal S}_2}{{\cal M}^4 r_0^3}\left[ \log \left(1-\frac{2 {\cal 
M}}{r_0}\right)\right] P_2(\theta)\,. \label{RRstarS2}
\end{align}
Using the same arguments as in the previous case, the current quadrupole moment of a 
soft ECO must go to zero as
\begin{equation}
{\cal S}_2\sim 1/\log\delta\,, \label{S2delta}
\end{equation}
or faster, as $\delta\to 0$. This result contrasts with the previous scenario where at 
linear order there was no restriction on the $\ell=1$ current multipole moment (i.e., 
the spin). By analyzing the second-order corrections to the Kretschmann 
scalar, we find that the curvature diverges in the BH limit as $\delta^{-2}$:
\begin{widetext}
\begin{align}
\label{eq:KS_2^2}
&{\cal K}^{(2)}\sim-\frac{225 \left(\bar{\cal S}_2\right)^2 \sin^4\theta}{256 {\cal M}^2 
(r_0-2 
{\cal M})^2}+\ldots-\frac{\bar{\cal S}_2^2}{{\cal M}^4}\left(\frac{45 }{28} \left(7 
B_2(\delta)+5\right)P_2(\theta)+\frac{135}{56}  \left(147 B_4(\delta)+202 
\right)P_4(\theta)\right)\log\delta	\,,
\end{align}
\end{widetext}
where the ellipsis accounts for other regular second-order terms. Similarly to the 
previous 
cases, in the above equation we defined 
\begin{equation}\label{eq:factorS2}
\bar {\cal M}_2=B_2(\delta)\bar{\cal S}_2^2+{\cal O}(\bar{\cal 
S}_2^4)\,, \quad \bar {\cal M}_4=B_4(\delta)\bar{\cal S}_2^2+{\cal 
O}(\bar{\cal S}_2^4)\,.
\end{equation}
Therefore, in this case we find that the saturation of Eq.~\eqref{S2delta} is not enough 
to prevent the curvature to blow up in the BH limit: the current quadrupole ${\cal 
S}_2$ must vanish as
\begin{equation}\label{eq:vanishings2}
\bar{\cal S}_2\to d_2\delta\,,
\end{equation}
or faster, as $r_0\to 2 {\cal M}$.
A similar argument can be presented for any multipole moment with $\ell>1$ entering at 
the leading order. 

The last term on the right hand side of Eq.~\eqref{eq:KS_2^2}, together with 
Eq.~\eqref{eq:vanishings2} and the factorization~\eqref{eq:factorS2}, would impose that 
the multipole moments ${\cal M}_2$ and ${\cal M}_4$ should vanish at least 
logarithmically in the $\delta\to0$ limit. Nonetheless, by extending this analysis to 
third order it is easy to show that they also must go to zero linearly, 
 \begin{equation}
 \bar{\cal M}_2\to b_2\delta\,, \qquad \bar{\cal M}_4\to b_4\delta\,,
 \end{equation} 
or faster,
similarly to the behaviour found for ${\cal S}_2$. (In the above equation $b_i$ are again
dimensionless constants which depend on the model.) Therefore, in this case all multipole 
moments must vanish linearly or faster.

The above argument applies to any solution sourced by a single multipole moment with 
$\ell>1$, showing that multipolar deviations from Kerr sourced by multipole moments other 
than the spin must vanish linearly or faster as $\delta\to0$. This striking difference 
relative to the spin-induced case is due to the fact that, already to first order, the 
solution contains irregular terms that must be suppressed as in Eq.~\eqref{RRstarS2}.
Since the leading-order moments with $\ell>1$ should already vanish logarithmically to 
the leading order, they induce power-law scalings in the curvature to next-to-the-leading 
order. The net result is that all moments should vanish at least linearly in this case.
This does not happen in the spin-induced case, where the first-order solution is 
everywhere regular, $g_{t\varphi}=2{\cal J}\sin^2\theta/r$, and therefore induces only 
logarithmic corrections at higher orders.

The discussion above was inferred from particular solutions where only one multipole 
moment was present to the leading order. However, as previously discussed, when two or 
more multipole moments are present at the leading order, the higher-order multipole 
moments can also be induced by mixing terms. One particular case of interest is the 
vanishing of a multipole moment 
induced simultaneously by spin-spin couplings and by the couplings with higher 
multipole moments. A representative case of this scenario is the ${\cal M}_4$ moment in 
the $\left[{\cal J}{\cal M}_2\right]$-solution, see Eq.~\eqref{M4_JM2}.
Although we did not compute the solution up to sufficiently high orders in the 
perturbation expansion to check explicitly the coupling terms in the curvature, results 
from the lower orders and the selection rules imply that the first two terms in 
Eq.~\eqref{M4_JM2} must vanish linearly (or faster) as 
$\delta\to0$, whereas the last term must vanish logarithmically, $\alpha_0\sim 
1/\log\delta$. The net result is
\begin{equation}
 \bar{\cal M}_4\to a_4 \frac{\chi^4}{\log\delta}+ b_4 \delta\,,
\end{equation}
or faster, in agreement with Eq.~\eqref{conjectureM}.

Finally, for all solutions we have checked that if the \emph{nonspin-induced} multipole 
moments vanish linearly (or faster) as $\delta\to0$ and the \emph{spin-induced} multipole 
moments vanish logarithmically (or faster), then the whole solution is regular in the 
BH limit also at higher order in the perturbative expansion.

\section{A ``soft-hair theorem'' conjecture for ECOs}
Given the general structure of the field equations, we expect the results obtained in the 
previous section to be valid at any order in the small-multipole expansion and to any 
order of the multipolar truncation. Thus, it is natural to conjecture that, within 
general relativity, the multipole moments of any (axisymmetric) ultracompact 
object whose exterior spacetime is perturbatively close to a Schwarzschild 
metric (i.e., the multipole moments of a soft ECO) must satisfy Eqs.~\eqref{mmECOM} 
and \eqref{mmECOS}, with
\begin{align}
\label{conjecture2}
\delta {\cal M}_\ell\to a_\ell\frac{\chi^\ell}{\log\delta}{\cal M}^{\ell+1}\,, \quad 
\delta {\cal S}_\ell \to c_\ell\frac{\chi^\ell}{\log\delta}{\cal M}^{\ell+1}\,.
\end{align}
or faster, as $\delta\to0$, if the multipole moments are spin-induced. For all other 
types of multipole moments,
\begin{align}
\label{conjecture2b}
\delta {\cal M}_\ell\to b_\ell {{\cal M}^{\ell+1}}{\delta}\,, \quad \delta {\cal 
S}_\ell \to d_\ell{{\cal M}^{\ell+1}}{\delta}\,,
\end{align}
or faster, as $\delta\to0$. In the above 
expressions, $a_\ell$, $b_\ell$, $c_\ell$ and $d_\ell$ dimensionless 
numbers of order unity or smaller, the spin and compactness dependence has been 
completely factored out. 
Overall, although we cannot prove it to arbitrary order, our results support the 
conjecture that the multipolar corrections of a soft ECO in the BH limit should behave as 
in Eqs.~\eqref{conjectureM} and \eqref{conjectureS}.

In other words, \emph{the closer a soft ECO is to a BH, the less hairy it will be}, in 
the manner specified by \eqref{conjecture2} or \eqref{conjecture2b}. 
This property can be seen as a weaker version of Birkhoff's theorem: beyond spherical 
symmetry the external spacetime is different from Kerr, but the multipolar deviations 
must die off as the BH limit is approached. Furthermore, spin-induced multipolar 
deviations die off much more slowly than any other.
While --~simply by continuity arguments~--
it is not surprising that $\delta {\cal M}_\ell,\delta {\cal S}_\ell\to0$ in 
the BH limit, the logarithmic scaling with $\delta$ in the spin-induced case is 
important, since it can give rise to potentially detectable effects, as we briefly 
discuss in the next section.


\section{Discussion and extensions} \label{sec:conclusion}
We have constructed and analyzed a class of axisymmetric 
solutions to Einstein's vacuum equations describing small (but otherwise 
generic) multipolar deformations of a Schwarzschild geometry. 
Our solution does not require equatorial symmetry and includes 
deformations induced by moments other than the spin. In fact, it is the
most general axisymmetric quasi-Schwarzschild solution to Einstein's vacuum 
equations.

We have introduced a classification for ECOs with ``soft'' and 
``hard'' hair, respectively. The former are characterized by a curvature at 
the surface which is comparable to that at the corresponding BH horizon, whereas 
the latter can have much larger curvature due to putative high-energy 
effects.
Self-consistency of the perturbative approach requires that our general solution 
belongs to the soft ECO class.

For this family of solutions, we showed that all \emph{spin-induced}
deviations from the BH multipole moments have to vanish logarithmically (or faster) as a 
function of the compactness parameter $\delta$ [see Eq.~\eqref{def:delta1}] in the BH 
limit. 
This logarithmic scaling already appeared in a particular case of our generic 
solution, namely in slowly-rotating gravastars~\cite{Pani:2015tga}, and we showed here 
that it is more generic.
On the other hand, multipolar deviations 
which are \emph{nonspin induced} must vanish linearly (or faster) and are therefore more 
difficult to constrain.
Overall, the approach to the BH limit is summarized in Eqs.~\eqref{conjectureM} and 
\eqref{conjectureS}. This ``soft-hair theorem'' implies that the more compact a soft ECO 
is, the less hairy it will be, thus extending Birkhoff's theorem to the case of small 
deviations from spherical symmetry.
Saturation of this (either logarithmic or linear) decay yields an 
upper bound on the compactness of quasi-Schwarzschild ECOs.
Furthermore, since spin-induced multipolar corrections die off much more slowly than any 
other in the BH limit, they are the dominant corrections and those that might provide the 
most stringent bounds on the ECO compactness.

Models of ECOs with soft hair can be constrained with future 
observations~\cite{Cardoso:2016rao,Cardoso:2016oxy,Cardoso:2017cfl,
Cardoso:2017cqb,Cardoso:2017njb,Glampedakis:2017cgd,Maselli:2017cmm,Allahyari:2018cmg}. In 
particular, 
extreme-mass 
ratio inspirals~(EMRIs) detectable by the future space mission LISA can place a 
constrain $\delta {\cal M}_2/{\cal M}^3<10^{-4}$ on the quadrupole moment of the 
central supermassive object~\cite{Babak:2017tow}. When the aforementioned decay of hair 
is saturated, the quadrupolar deviation of soft ECOs reads
\begin{eqnarray}
 \frac{\delta{\cal M}_2}{{\cal M}^3}\sim\left\{\begin{array}{l}
                                        |\log\delta|^{-1}\sim 
10^{-2}\left|\log\left(\frac{10^6 M_\odot}{\ell_P}\right)\right|^{-1}\\
\delta^{-1} \sim 10^{-4} \left(\frac{L}{10^6 M_\odot}\right)
                                       \end{array}
\right.\,,
\end{eqnarray}
for spin- and nonspin-induced quadrupolar hair, respectively. In the former case we 
defined $\delta=\ell_P/(2{\cal M})$, where $\ell_P$ 
is the Planck length, whereas in the latter case $\delta=L/(2{\cal M})$, where $L\approx 
50\,{\rm km}$. In 
other words, an EMRI detection by LISA has the potential to constrain spin-induced 
multipolar deviations from Kerr even for objects motivated by quantum-gravity 
considerations~\cite{Mazur:2004fk,Mathur:2005zp,Mathur:2008nj,Barcelo:2015noa, 
Danielsson:2017riq,Berthiere:2017tms}, whereas multipolar deviations which are not spin 
induced can be constrained only for objects with compactness smaller than $M/r_0 \approx 
0.49995$, nonetheless an impressive constraint. The scaling 
rules~\eqref{conjectureM}~and~\eqref{conjectureS} imply that a quadrupole 
moment measurement of a soft ECO will always be dominated by the 
spin-induced contribution, unless
\begin{equation}
 \chi\ll \sqrt{\delta|\log\delta|}\,,
\end{equation}
where we assumed $b_2/a_2\sim{\cal O}(1)$. The above upper bound is very small for 
realistic values of $\delta$, e.g. $\chi\ll 0.03$ when $\delta\approx10^{-4}$.

A detailed studied on the phenomenology of EMRIs in the case of ECOs with soft hair and 
the connection between our solution and existing parametrizations (e.g. with ``bumpy'' 
BHs~\cite{Collins:2004ex,Vigeland:2009pr}) will appear elsewhere.

Our argument is based on the properties of the external spacetime and is 
therefore independent of the internal structure of the body. There are, however, 
some caveats that are worth discussing:
\begin{itemize}
 \item We assumed vacuum Einstein equations, so strictly speaking our analysis is not 
valid in modified gravity or in the presence of long-range fields in general relativity. 
However, we expect our argument to remain qualitatively valid if putative 
external fields die off sufficiently fast (e.g., for boson stars, in particular 
the most compact ones, and for BHs in effective-field-theory 
corrections to general 
relativity
\cite{Mignemi:1992nt,Pani:2011gy,Yunes:2011we,Sotiriou:2013qea,Cardoso:2018ptl}, which 
are suppressed at large distance by 
high-curvature terms). 
In particular, our parametrization could approximately describe the external metric of 
spinning BHs without equatorial symmetry which were recently constructed in extensions of 
general relativity~\cite{Cardoso:2018ptl} and with specific matter 
sources~\cite{Cunha:2018uzc}.
We also expect that an extension to Einstein-Maxwell 
should be relatively straightforward and qualitatively similar to the vacuum 
case.
 \item We assumed that the 
multipole moments are perturbatively small. While this necessarily restricts the 
analysis to soft ECO models, it might not always be the case. For example, the angular 
momentum of a spinning boson star is quantized so its spin-induced 
multipole moments cannot be made arbitrarily small. However, stable boson stars 
have a maximum compactness that is not continuously connected to that of a BH, so they 
are outside the scope of our study. 
\end{itemize}

An important open issue concerns the stability of these geometries. The answer to this problem depends on the 
internal composition of the object or, equivalently, on the boundary conditions for time-dependent perturbations at the 
surface, which we left unspecified (also in the axisymmetric, stationary case) for the sake of clarity. As such, the 
stability issue can be assessed only case by case. Some particular examples of our general solution (namely, gravastars) 
are linearly (mode) stable under radial~\cite{Visser:2003ge} and nonradial~\cite{Pani:2009ss} perturbations for 
densities below that corresponding to the maximum mass; in this respect they are therefore similar to ordinary neutron 
stars.
On the other hand, linearized gravitational fluctuations of static ultracompact horizonless objects with perfectly 
reflective boundary conditions are extremely long-lived and decay no faster than 
logarithmically~\cite{Keir:2014oka}. The long damping time of these modes 
has led to the conjecture that these objects might be nonlinearly unstable~\cite{Keir:2014oka,Cardoso:2014sna}, although 
most likely this conclusion and the putative endstate of the instability depend on the specific model.
Likewise, \emph{spinning} ultracompact horizonless objects with perfectly reflective boundary conditions are 
\emph{linearly} unstable toward the ergoregion instability~\cite{Friedman:1978wla,Moschidis:2016zjy,Cardoso:2007az} 
(see~\cite{Brito:2015oca} for a review).
Also in this case the instability depends on the specific model. In particular, partial absorption in the interior 
might quench the instability completely~\cite{Maggio:2017ivp,Maggio:2018ivz}.

A possible extension of our analysis concerns relaxing the assumption of soft hair. In 
axisymmetry, this can be done, even at the nonlinear level, using the 
integrability of stationary axisymmetric metrics (see e.g. 
Refs.~\cite{1997PhLA..230....7B,1984PhLA..103..374T}). For static (i.e., without  angular 
momentum) deformations, these solutions belong to the 
Weyl class~\cite{Weyl}.

A discussion of the no-hair properties of ECO models with hard hair will appear 
elsewhere~\cite{inprep}. Future work will also 
include a study of the geodesic properties of the quasi-Schwarzschild solution in 
the general case and of the detectability of multipolar corrections with current and 
future observations.

Finally, although our main motivation was to study ECOs, the family of perturbative 
solutions constructed in this work might also serve to study generic (axisymmetric) 
deformations of a neutron star, for example those sourced by an intrinsic quadrupole 
moment.

\begin{acknowledgments}
We are indebted to George Pappas for interesting discussion and comments on a draft 
manuscript and to Kento Yagi for relevant correspondence.
GR acknowledges the kind hospitality of Johns Hopkins University where this work 
has been finalized.
PP acknowledges financial support provided under the European Union's H2020 ERC, Starting 
Grant agreement no.~DarkGRA--757480. RE is supported by ERC Advanced Grant GravBHs-692951 
and MEC grant FPA2016-76005-C2-2-P.
This project has received funding from the European Union's Horizon 2020 research and 
innovation programme under the Marie Sklodowska-Curie grant agreement No 690904.
The authors would like to acknowledge networking support by the COST Action CA16104 and 
support from the Amaldi Research Center funded by the MIUR program "Dipartimento di 
Eccellenza" (CUP: B81I18001170001).
\end{acknowledgments}
%

\bibliographystyle{utphys}
\bibliography{refs}	
\end{document}